\documentclass{aa}
\usepackage{graphicx}
\usepackage[varg]{txfonts}
\usepackage{units}
\usepackage{mathtools}
\usepackage{color}
\usepackage{natbib,twoopt}
\usepackage[breaklinks=true]{hyperref} 
\bibpunct{(}{)}{;}{a}{}{,}             
\makeatletter
  \newcommandtwoopt{\citeads}[3][][]{\href{http://adsabs.harvard.edu/abs/#3}%
    {\def\hyper@linkstart##1##2{}%
     \let\hyper@linkend\@empty\citealp[#1][#2]{#3}}}
  \newcommandtwoopt{\citepads}[3][][]{\href{http://adsabs.harvard.edu/abs/#3}%
    {\def\hyper@linkstart##1##2{}%
     \let\hyper@linkend\@empty\citep[#1][#2]{#3}}}
  \newcommandtwoopt{\citetads}[3][][]{\href{http://adsabs.harvard.edu/abs/#3}%
    {\def\hyper@linkstart##1##2{}%
     \let\hyper@linkend\@empty\citet[#1][#2]{#3}}}
  \newcommandtwoopt{\citeyearads}[3][][]%
    {\href{http://adsabs.harvard.edu/abs/#3}
    {\def\hyper@linkstart##1##2{}%
     \let\hyper@linkend\@empty\citeyear[#1][#2]{#3}}}
  \renewcommand*\aa@pageof{, page \thepage{} of \pageref*{LastPage}}
\makeatother
\usepackage{hyperref}
\hypersetup{pdfpagemode = {UseNone},
            pdftitle = {Planet-disc interaction in laminar and turbulent discs},
            pdfcreator = {\LaTeX},
            pdfproducer = {pdfeTeX-0.\the\pdftexversion\pdftexrevision},
            pdfauthor = {Moritz Stoll, Giovanni Picogna, Wilhelm Kley},
            pdfsubject = {},
            pdfview = {FitH},
            pdfstartview = {FitH},
            colorlinks = {true},
            linkcolor = [rgb]{0,0.35,0.7},
            citecolor = [rgb]{0,0.35,0.7},
            filecolor = [rgb]{0.61,0,0},
            urlcolor = [rgb]{0,0.35,0.7},
           }

\usepackage{afterpage}
\usepackage{silence}
\newcommand{\beq}{\begin{equation}}
\newcommand{\eeq}{\end{equation}}

\begin{document}

\title{Planet-disc interaction in laminar and turbulent discs}

\author{Moritz~H.~R.~Stoll\inst{1} \and
Giovanni Picogna\inst{1,2} \and
Wilhelm Kley\inst{1}}

\institute{
Institut f\"{u}r Astronomie und Astrophysik, Universit\"{a}t T\"{u}bingen,
Auf der Morgenstelle 10, D-72076 T\"{u}bingen, Germany\\
\email{\{moritz.stoll@, wilhelm.kley@\}uni-tuebingen.de}
\and
Universit\"{a}ts-Sternwarte, Ludwig-Maximilians-Universit\"{a}t München,
Scheinerstr. 1, D-81679 M\"{u}nchen, Germany\\
\email{picogna@usm.lmu.de}
}

\date{Received February, 21, 2017 / Accepted \today}

\abstract{In weakly ionized discs turbulence can be generated through the vertical shear instability (VSI).
Embedded planets feel a stochastic component in the torques acting on them which can impact their migration.
In this work we study the interplay between a growing planet embedded in a protoplanetary disc and the VSI-turbulence.
We performed a series of three-dimensional hydrodynamical simulations for locally isothermal discs
with embedded planets in the mass range from 5 to 100 Earth masses.
We study planets embedded in an inviscid disc that is VSI unstable, becomes turbulent and 
generates angular momentum transport with an effective $\alpha = 5 \cdot 10^{-4}$.
This is compared to the corresponding viscous disc using exactly this $\alpha$-value.

In general we find that the planets have only a weak impact on the disc turbulence.
Only for the largest planet ($100 M_\oplus$) the turbulent activity becomes enhanced inside of the planet.
The depth and width of a gap created by the more massive planets ($30, 100 M_\oplus$) in the turbulent
disc equal exactly that of the corresponding viscous case, leading to very similar torque strengths acting on the
planet, with small stochastic fluctuations for the VSI disc. At the gap edges vortices are generated that
are stronger and longer lived in the VSI disc.
Low mass planets (with $M_p \leq 10 M_\oplus$) do not open gaps in the disc in both cases but generate
for the turbulent disc an over-density behind the planet that exerts a significant negative torque.
This can boost the inward migration in VSI turbulent discs well above the Type I rate.

Due to the finite turbulence level in realistic three-dimensional discs the gap depth will always be limited
and migration will not stall in inviscid discs.
}


\maketitle

\section{Introduction}

The longstanding problem of identifying the mechanism responsible for transporting angular momentum in accretion discs and explain their observed lifetimes has been tackled since the early works of \citetads{1973A&A....24..337S} and \citetads{1974MNRAS.168..603L}.
The physical process that has been identified having an important contribution is the magnetorotational instability (MRI) (\citeads{1991ApJ...376..214B}; \citeads{1991ApJ...376..223H}). However, within the context of planet formation, the basic requirement of well ionised media to trigger the MRI is generally not fulfilled in protoplanetary discs \citepads{1996ApJ...457..355G}.

Several purely hydrodynamical instabilities have been proposed to drive the angular momentum, even in the non-ionised regions of the disc (also referred to as "dead zones"), that  can develop under special conditions \citepads[see e.g.][and references therein]{2013MNRAS.435.2610N}. One of the mechanisms that are more generally applicable is the vertical shear instability (VSI) which was first discovered in the context of differentially rotating stars (\citeads{1967ApJ...150..571G}; \citeads{1968ZA.....68..317F}) and then later applied also to accretion discs \citepads{1998MNRAS.294..399U}. More recently, the VSI has been modelled by direct numerical simulations and proven effective in locally isothermal discs (\citeads{2013MNRAS.435.2610N}) as well as in fully radiative discs \citepads{2014A&A...572A..77S}.
Within the context of the planet formation process, these instabilities can strongly impact the dust dynamics, by increasing for example the dust concentration inside vortices (\citeads{1995A&A...295L...1B}; \citeads{2006ApJ...639..432K,2016MNRAS.458.3927B}) and regulating its settling and growth \citepads{2005A&A...434..971D}.
In particular, \citetads{2016A&A...594A..57S} studied the effect of the VSI on the dust dynamics and find a strong clustering of dust particles by the large-scale vertical motion induced by the instability.
At the same time, the relative speed between individual particles is significantly reduced which can be highly beneficial for the planet formation process. Furthermore, a strong outward migration of sub-mm size dust is observed in the upper layers of the disc that can replenish the outer disc regions of solid material.

An embedded protoplanet can strongly affect the disc structure in its vicinity by creating density waves departing from its location and by depleting the co-orbital region. These asymmetries in the disc's density generate a gravitational torque acting onto the planet and force it to migrate through the disc \citepads[see e.g.][and references therein]{2012ARA&A..50..211K}. The direction and magnitude of this migration depends strongly on the local properties (viscosity and thermodynamics) of the disc.
The VSI generates strong vertical motions that can perturb the disc equilibrium,
potentially affecting the resulting torque acting onto the planet.
Previously, direct simulations of planets embedded in turbulent discs have been performed only for the magnetic case of MRI unstable discs \citep{2004MNRAS.350..849N,2005A&A...443.1067N,2011ApJ...736...85U}.

In this work we focus on the signatures that are induced by a planet in a protoplanetary disc in which the VSI is operating, and calculate the
torque acting on the planet. We compare our results directly to the classical viscous $\alpha$-disc models in order to study
the main differences that this simpler prescription can or cannot reproduce.
Additionally, we compare the strength of the torques acting on the planet in hydrodynamic VSI-turbulence to that generated in MRI-turbulence.
The paper is organised as follows. In Sec.~\ref{sec:setup} we describe the numerical method used to model the disc, the planet and the parameter space studied. In Sec.~\ref{sec:results} we outline the results of our analysis, and in Sec.~\ref{sec:discussions} we discuss our major findings and draw the conclusions.

\section{Setup}\label{sec:setup}

We used the \textsc{pluto} code \citepads{2007ApJS..170..228M} to model a three-dimensional section of a locally isothermal accretion disc
in spherical coordinates (r, $\theta$, $\phi$) with embedded planets of different masses. 
To study the planet-disc interaction we compared two sets of simulations. In the first series we ran inviscid models that generated 
direct turbulence through the VSI and in the second series we ran viscous models using an effective $\alpha$-value adapted to that of the 
VSI turbulent disc. 
Radially, the disc section extends from $\unit[0.4]{r_\mathrm{p}}$ to $\unit[2.5]{r_\mathrm{p}}$, 
where $r_\mathrm{p} = \unit[5.2]{au}$ is the planet position. 
\citet{Lin2015ApJ...811...17L} estimated that the VSI for a minimum mass solar nebula would operate in the range between 
$5-50$ au, so we are just at the inner boundary of their suggested range. In our situation planets have already formed and considering a later phase
of the planet evolution process the surface density in solids is reduced which would make the VSI more efficient
due to the enhanced cooling. At the same time, the configuration we study can be considered an exemplary case, centered at 5.2 au, 
that could be scaled easily to other regimes due to the locally isothermal assumption.
In the vertical direction the models cover 5 scale heights above and below the equator, and a full annulus in the azimuthal direction.
This domain is covered for the standard model by $N_r=600, N_\theta=128$, and $N_\phi = 1024$ gridcells that are spaced logarithmically
in the radial and equidistant in the other directions. 
At the location of the planet ($r=1$ in our units) the radial size of a gridcell is $\Delta r = 0.003$,
and the ratio of gridsizes near the location of the planet is given by $\Delta r:r \Delta \theta: r \Delta \phi = 1.0:1.28:2.012$.
A comparison model with double spatial resolution is discussed in section \ref{subsec:double}.
The main parameters of the simulations are summarised in Tab.~\ref{Tab:sum}.

\subsection{Gas component}\label{par:gasdisc}
The initial disc is axisymmetric and, even though we use spherical coordinates in the simulations,
we state the initial conditions in cylindrical coordinates $(R,Z,\phi$).
The initial density profile created by force equilibrium is given by
\begin{equation}
    \rho(R,Z)=\rho_0\, \left(\frac{R}{R_\mathrm{p}}\right)^{p}\, \exp{\left[\frac{G M_\mathrm{s}}{c_\mathrm{s}^2}\left(\frac{1}{r}-\frac{1}{R}\right)\right]}\,,
    \label{eq:surfprof}
\end{equation}
where $\rho_0$ is the gas mid-plane density at $R = R_\mathrm{p} = r_\mathrm{p}$, $p=-1.5$ the density exponent, and $c_\mathrm{s}$ the isothermal sound speed. 
In our simulations we use $\rho_0 = 2.07 \cdot 10^{-11}$ g/cm$^3$ which translates into $\Sigma(R_0) = 200$ g/cm$^2$
for our chosen disc thickness.
The disc is modelled with a locally isothermal equation of state, which keeps the initial temperature stratification fixed throughout the whole simulation. We assume a constant aspect ratio $H/R = 0.05$, which corresponds to a temperature profile
\begin{equation}
  T(R) = T_0\, \left(\frac{R}{R_\mathrm{p}}\right)^{q} \,,
  \label{eq:tempprof}
\end{equation}
with $q=-1$ and $T_0 = \unit[121]{K}$.

The gas moves with an azimuthal velocity given by the Keplerian speed around a $\unit[1]{M_\sun}$ star, corrected by the pressure support \citepads{2013MNRAS.435.2610N}
\begin{equation}
  \Omega(R,Z) = \Omega_\mathrm{K} \left[ (p +q)\left( \frac{H}{R} \right)^2 + (1+q) - \frac{qR}{\sqrt{R^2 + Z^2}} \right]^{\frac{1}{2}} \,.
  \label{eq:omega}
\end{equation}

At the inner  and outer  boundary we damp the density, radial and vertical velocity to the initial values with a timescale of half a local orbit \citepads[see e.g.][]{2006MNRAS.370..529D}, in order to reduce numerical issues due to the interaction of the VSI with the boundary and to prevent reflection of the spiral wave at the boundaries. The damping is applied in the intervals \unit[[0.4,0.5]]{$r_\mathrm{p}$} and \unit[[2.3,2.5]]{$r_\mathrm{p}$}.
For the kinematic viscosity in the $\alpha$-disc model we use $\nu = 2/3 \alpha c_\mathrm{s} H$, and adopt a constant $\alpha$ parameter of $\alpha = 5 \cdot 10^{-4}$, which is close to the value obtained from the VSI model (see Fig.~\ref{fig:alpha}).
To bring the VSI model into equilibrium, we evolve it for $\unit[200]{orbits}$ before embedding the planet.

\begin{table}[tb]
    \caption{Model parameter}
    \label{Tab:sum}
    \centering
    \begin{tabular}{c c }
        \hline\hline
        Parameter & model \\
        \hline
        Radial range [$5.2\ \mbox{au}$] & $0.4$ - $2.5$  \\
        Vertical range [H] & $\pm 5$  \\
        Phi range [rad] & $0$ - $2\ \pi$   \\
        Radial grid size & $600$  \\
        Theta grid size & $128$  \\
        Phi grid size & $1024$   \\
        Planet mass [$M_\oplus$] & $5$, $10$, $30$, $100$   \\
        \hline
    \end{tabular}
\end{table}
\subsection{The planet}\label{par:planet}
We embed a planet that orbits its parent star on a circular orbit with a
radius $r_\mathrm{p}=1$ and a mass in the range $\unit[[5, 10, 30, 100]]{M_\oplus}$.
These 4 planets have the following Hill radii, $\unit[[0.017,0.022,0.031, 0.045]]{R_\mathrm{p}}$.
Using the grid resolution for the standard model in Table \ref{Tab:sum} this implies that there are about 100 gridcells within the Hill radius for
the smallest planet mass and 740 cells for the largest planet mass.
Gas accretion is not allowed, and the planet does not feel the disc,
so its orbital parameters remain fixed during the whole simulation. The gravitational potential of planet is implemented with a cubic expansion in the vicinity of the planet location \citepads{2006A&A...445..747K}
\begin{equation}
  \Phi_\mathrm{p} =
  \begin{dcases}
    - \frac{GM_\mathrm{p}}{d} \left[\left( \frac{d}{d_\mathrm{rsm}} \right)^4- 2 \left( \frac{d}{d_\mathrm{rsm}} \right)^3 + 2 \left( \frac{d}{d_\mathrm{rsm}} \right) \right],& \text{for } d < d_\mathrm{rsm}\\
    - \frac{GM_\mathrm{p}}{d},& \text{for } d \geq d_\mathrm{rsm},
  \end{dcases}
  \label{potential}
\end{equation}
where $d$ is the distance of a gas element from the planet location, and the smoothing length $d_\mathrm{rsm}=\unit[0.5]{R_\mathrm{Hill}}$ is adopted in order to avoid singularities.
To ensure a smooth start during the initial phase of the simulations, the planetary mass is increased slowly over the first $\unit[20]{orbits}$ of the simulation. We run the simulations for $200$ orbital periods of the planet.


\section{Results}\label{sec:results}

\subsection{The effective viscosity of the disc with embedded planets}
For the VSI models the $\alpha$-parameter is not set by the viscosity prescription. Instead, the VSI generates eddies that transport angular momentum self-consistently. We can then calculate the efficiency of this process by evaluating an effective $\alpha$-parameter through the Reynolds stress resulting from the turbulence.
We calculate the Reynolds stress in cylindrical coordinates $(R, Z, \phi$)
\begin{equation}
    R_{R,\phi} = \rho u_R \delta u_{\phi}  \,,
    \label{eq:Rey}
\end{equation}
where $\delta u_{\phi}$ is the local difference from the equilibrium azimuthal velocity, which we calculate by time averaging $u_{\phi}$ for each grid cell from orbit $60$ to orbit $200$ over $70$ snap shots.

We then calculate the dimensionless $\alpha$-parameter as a function of the radius,
\begin{equation}
    \alpha(r) = \frac{\left< R_{R,\phi} \right>_{t,\theta,\phi}}{\left< P \right>_{t,\theta,\phi}} \,,
\end{equation}
where $P$ is the pressure and $\left< \right>_{t,\theta,\phi}$ denotes the average over time ($\unit[140]{orbits}$) and the whole vertical and azimuthal domain.

\begin{figure}[tb]
    \centering
    \includegraphics{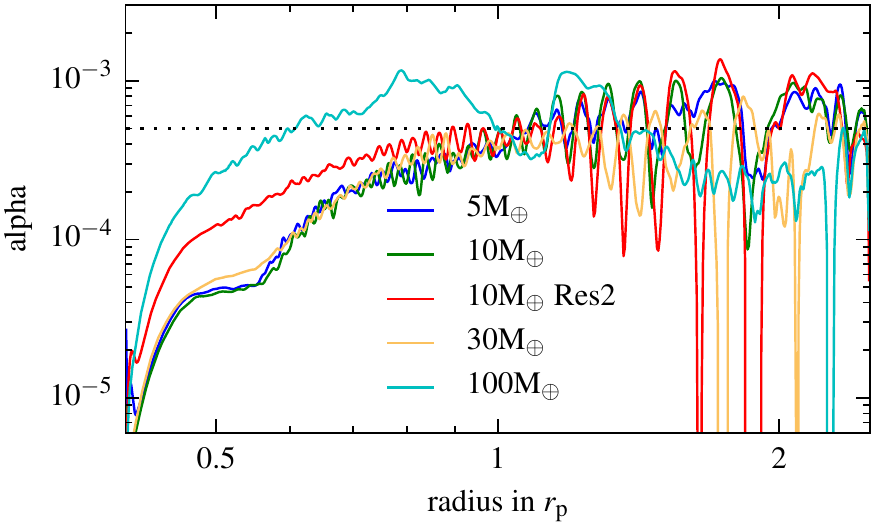}
    \caption{Comparison of the $\alpha$-parameter for our VSI-active simulations with embedded planets of different mass.
    The planet is located at  $r = r_\mathrm{p}$. 'Res2' indicates a model with double resolution. The dotted line is at $\alpha = 5 \cdot 10^{-4}$.}
    \label{fig:alpha}
\end{figure}

\begin{figure*}[t!]
    \centering
    \includegraphics{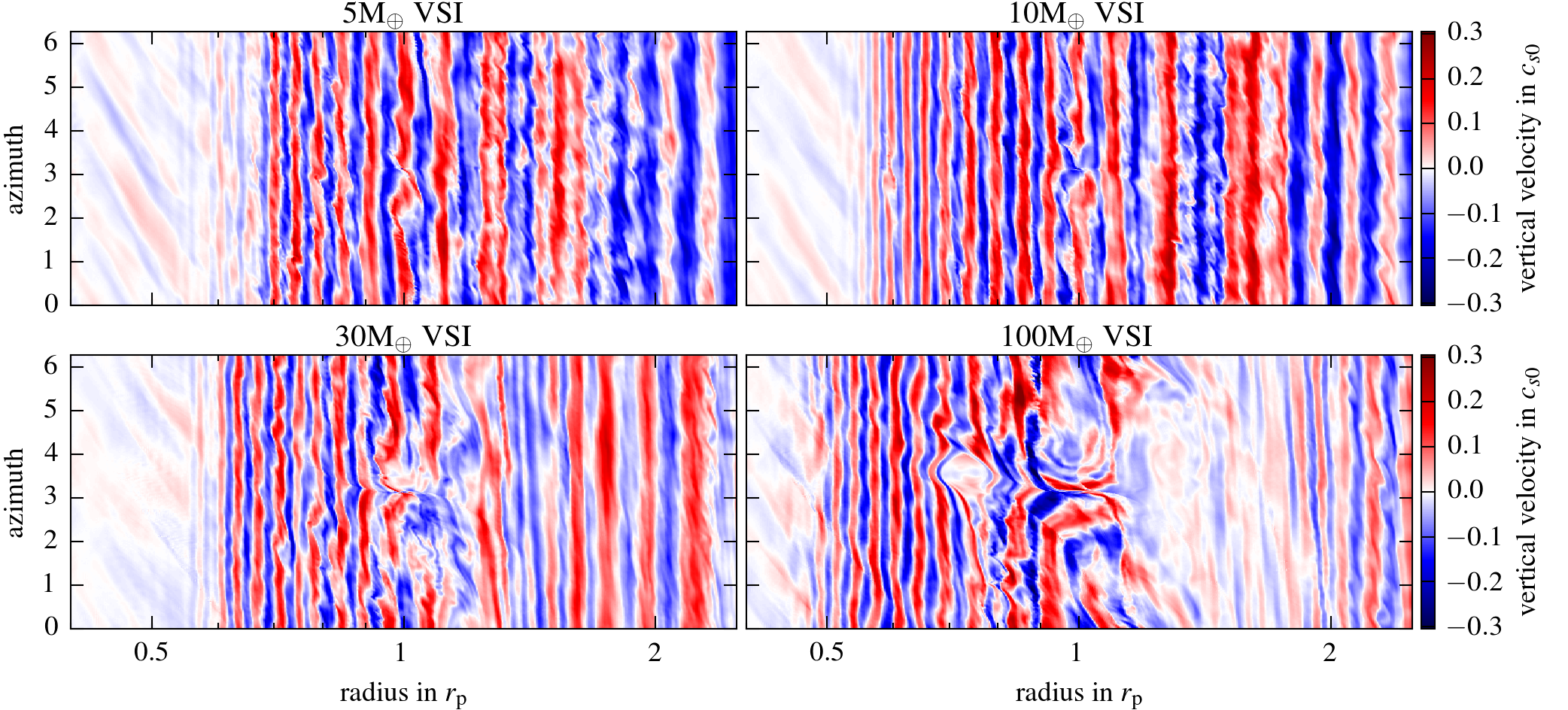}
    \caption{Vertical velocity in the midplane of the VSI-disc after $\unit[198]{orbits}$ with embedded planets of different mass. Red colour denotes an upward motion and blue downwards. Larger planets clearly disrupt the usual VSI-modes close to the planet. In the last panel (\unit[100]{$M_\oplus$} planet) vortices are visible at both sides of the planet.
Also visible in the last panel is the more extended VSI activity near to the inner boundary compared to the other panels.}
    \label{fig:verticalVelocity}
\end{figure*}

In Fig. \ref{fig:alpha} we compare the $\alpha$-parameter for our simulations of VSI-active discs with embedded planets of different mass. The inner damping region from \unit[0.4]{$r_\mathrm{p}$} to \unit[0.5]{$r_\mathrm{p}$} is very effective in suppressing the VSI, which is intended to reduce the interaction with the boundary. The wave killing zone is slightly less effective in suppressing the VSI for an increased grid resolution close to the inner boundary (model 'Res2' for the 10 $M_\oplus$ model), while there is no difference in the outer disc. Interestingly, the VSI is more resilient to the wave killing zone for the \unit[100]{$M_\oplus$} case. There the VSI is stronger and can access a region much closer to the inner boundary. This behaviour is also highlighted in Fig.~\ref{fig:verticalVelocity}, where the vertical velocity in the midplane of the disc (induced by the VSI) reaches $0.5 c_\mathrm{s0}$ for the more massive case.
One possibility for this extended turbulent activity might by the action of the spiral wave instability (SWI)
where the presence of a massive planet can excite unstable inertial-gravity waves \citep{2016ApJ...829...13B,2016ApJ...833..126B}
which could enhance the turbulent level above that of the VSI alone.

In the top panels of Fig.~\ref{fig:verticalVelocity} we see that small planetary masses are not able to influence the strength of the VSI, even close to their location. The strong deviation in the $\alpha$-profile seen in Fig.~\ref{fig:alpha} for the large mass planets can be directly linked to the bottom panels of Fig.~\ref{fig:verticalVelocity} with the formation of strong vortices at the gap edges carved by the planets. The vortex at the outer edge is strong enough to suppress the VSI in the region \unit[[1.2, 2.0]]{$r_\mathrm{p}$}. For the \unit[30]{$M_\oplus$} case, the variation of the surface density profile is sufficient to reduce the effectiveness of the VSI in close proximity to the planet.

Finally, in the outer region, beginning at \unit[2]{$r_\mathrm{p}$}, the VSI can operate unaffected by the planet and leads to an $\alpha$-parameter close to $10^{-3}$, which is approximately the value inferred from observations \citepads[see e.g.][]{2009ApJ...700.1502A}. Only for the largest planet the VSI is slightly suppressed due to the formation of a large vortex outside of the planetary orbit.

Altogether, we can see that in these simulations the overall amplitude and radial dependence of $\alpha$ is very similar to our previous runs,
see for example Fig.~3 in \citetads{2014A&A...572A..77S}. The fact that $\alpha(r)$ drops towards smaller radii appears to be the result of 
two effects. First, the inner damping zone reduces the strength of the turbulence, second the spatial resolution of the models
may not be sufficient as the result for a model with double resolution (shown for the 10 $M_\oplus$ planet and marked 'Res2' in Fig.~\ref{fig:alpha}) 
indicates. More discussion on the spatial resolution will be given in section \ref{subsec:double} below.

\subsection{Vortices and vorticity}\label{sec:vortex}

\begin{figure*}[t!]
    \centering
    \includegraphics{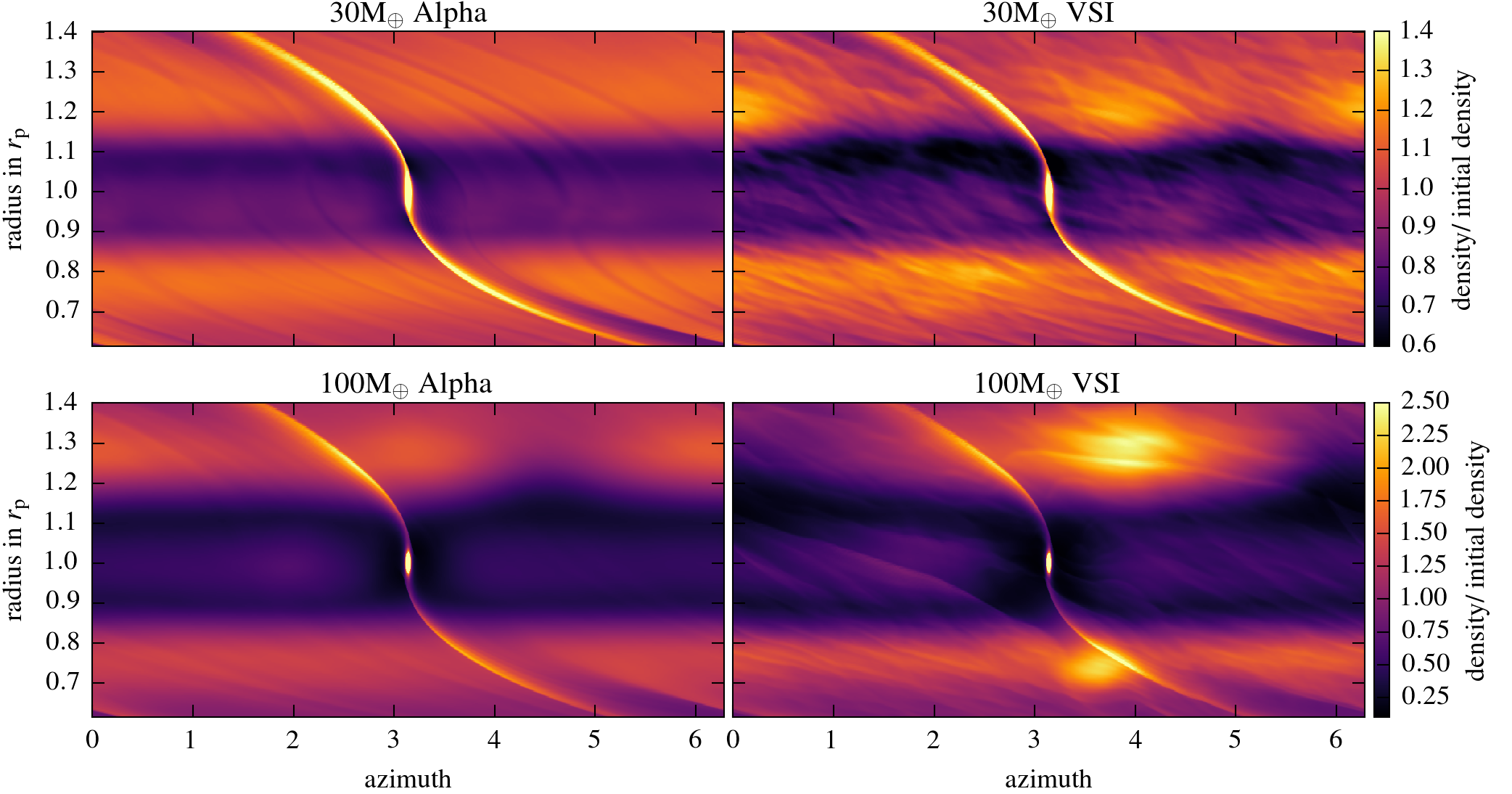}
    \caption{Density in the midplane relative to the initial density after $\unit[198]{orbits}$ for the inviscid VSI-active simulations and
   viscous disc models using $\alpha = 5 \cdot 10^{-4}$.
   {\bf Upper Panels}: after $\unit[198]{orbits}$ for $M_\mathrm{p}=\unit[30]{M_\oplus}$. {\bf Lower Panels}: same for $M_\mathrm{p} = \unit[100]{M_\oplus}$.}
    \label{fig:vortex}
\end{figure*}

Vortex creation by embedded planets is a phenomenon occurring in nearly inviscid discs \citep{2003ApJ...596L..91K,Artymowicz2007A&A...471.1043D},
while they are usually suppressed by viscosity over time \citep{2006MNRAS.370..529D}.
It is thus unclear what happens in VSI simulations that are able to generate a viscosity on the order of $\alpha = 5 \cdot 10^{-4}$. In fact, in the simulations with VSI and $M_\mathrm{p} = \unit[100]{M_\oplus}$ we see multiple vortices developing at the outer edge of the gap after the slow introduction of the planet (over $\unit[20]{orbits}$). They merge to a single vortex over the next $\unit[100]{orbits}$, which then remains stable, and even increases slightly in strength up to the end of the simulation.
This vortex, near the end of the simulation\footnote{At $\unit[200]{orbits}$ the vortex sits at the periodic boundary, thus we show one earlier time slice, where it is in the middle of the domain.}, is shown in Fig. \ref{fig:vortex} on the right side of the lower panels. There we not only can see a vortex at the outer gap edge, but also a smaller vortex that has formed at the inner edge of the gap. Both vortices exist over the whole simulated timescale of $\unit[200]{orbits}$. This is in contrast with the $\alpha$-disc on the left side of the lower panel, where we can still see two weak vortices that are dissipating in the long run due to the effect of viscosity.

Now we compare directly the inviscid VSI-active simulations with the results of a viscous disc using $\alpha = 5 \cdot 10^{-4}$. The upper panel of Fig. \ref{fig:vortex} shows the simulations with $M_\mathrm{p} = \unit[30]{M_\oplus}$. As expected, the $\alpha$-viscosity disc does not show large scale vortices, but the VSI simulations again introduces a smaller vortex, which is also stable over the runtime of the simulation. The runs with smaller planets do not have a gap and thus do not show large scale vortices in this region, but instead smaller vortices are visible.

To further analyse the vortices we calculate a normalised vorticity
\begin{equation}
    \overline{\vec{\omega}}_{\theta} = \left( \frac{r}{r_{\mathrm{p}}} \right)^{3/2}\left( \nabla \times \vec{u} \right)_{\theta} \,.
    \label{eq:vort}
\end{equation}
For undisturbed keplerian rotation this leads to $\overline{\vec{\omega}}_{\theta} = \vec{\omega}_{\theta K}$ independent of radius, where $\vec{\omega}_{\theta K}$ is the vorticity at $r=r_\mathrm{p}$.

\begin{figure*}[t!]
    \centering
    \includegraphics{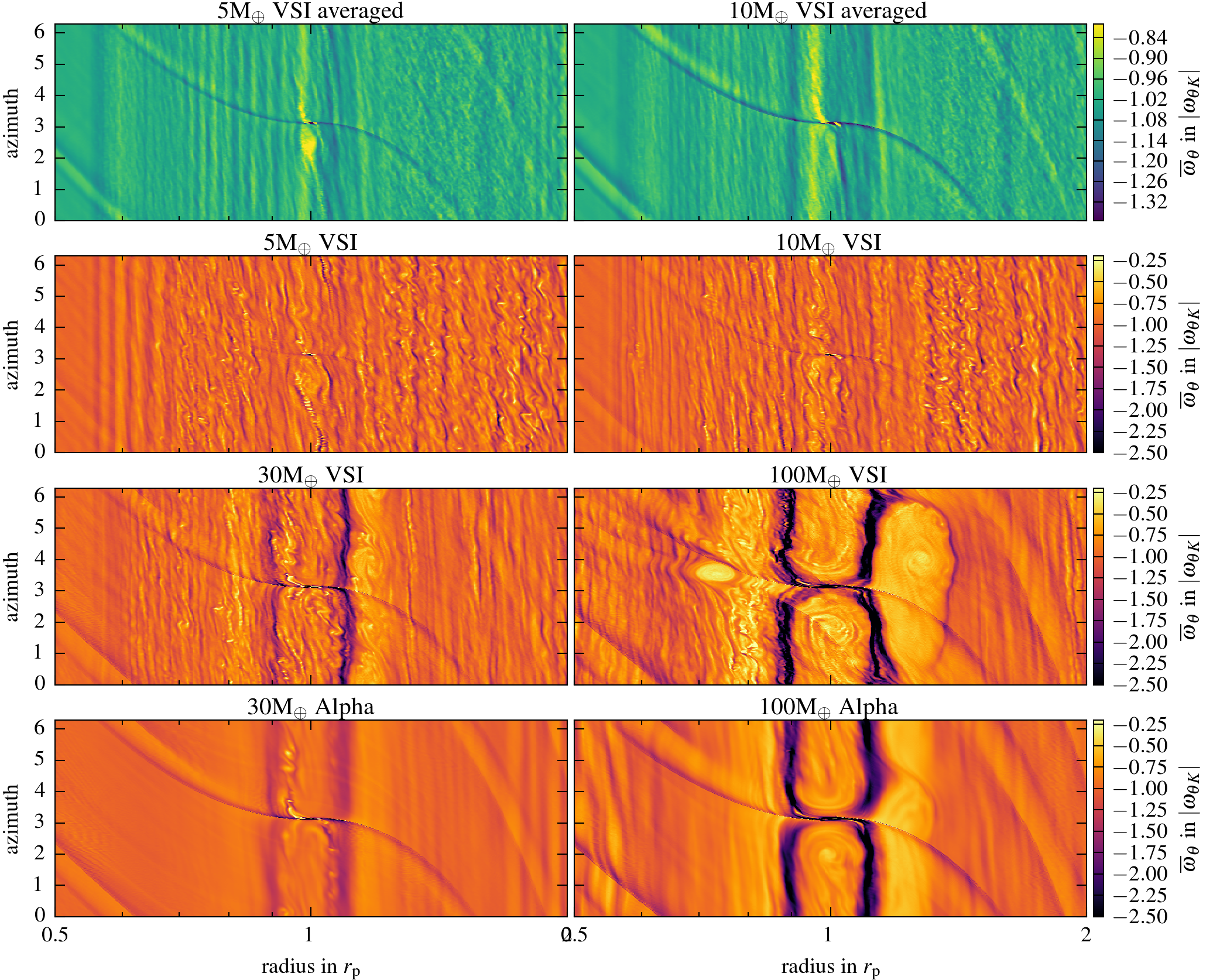}
    \caption{{\bf First Row:}  Normalised vorticity (eq.~\ref{eq:vort}) in the disc midplane averaged over the last $\unit[100]{orbits}$ in the VSI disc. The increased vorticity close to the planet is clearly visible. {\bf Second and Third Row}: Normalised vorticity in the midplane for the VSI disc after $\unit[198]{orbits}$. {\bf Last Row}: Normalised vorticity in the midplane for the viscous $\alpha$-disc after $\unit[198]{orbits}$.}
    \label{fig:vorticity2D}
\end{figure*}

The normalised vorticity in the disc's midplane is displayed in Fig. \ref{fig:vorticity2D} for various models.
The second and third row show the normalised vorticity for the different planets in the VSI discs after $\unit[200]{orbits}$. 
The creation of small vortices in regions with active VSI turbulence was already noticed and explained by \citetads{2016MNRAS.456.3571R}.
They are only short lived in locally isothermal discs \citepads{2016MNRAS.456.3571R} and are only weakly influenced by the lower
mass planets in the main part of the disc.
However, in its close vicinity the planet impacts the dynamics of the vortices, as shown in the first row where
the vorticity of the two low mass planets is plotted averaged in time over 50 snapshots, beginning with orbit 100 and ending with orbit 200. 
We can see that the characteristic horseshoe motion of the flow transports the vorticity created close to the planet, 
which leads to an increase in the surface density along the inner horseshoe orbit. 
For the smallest planet with $\unit[5]{M_\oplus}$ this leads, on average, to an region of enhanced vorticity in close proximity behind the planet (top left panel in Fig.~\ref{fig:vorticity2D}),
which affects the torque on the planet as we will see in Sec.~\ref{sec:torques}.
For the $\unit[10]{M_\oplus}$ planet the effect becomes somewhat weaker but there is still an enhanced vorticity visible just inside
of the planetary orbit spread out in azimuth.

For the larger planet masses with $30$ and $\unit[100]{M_\oplus}$ the presence of the planet leads to the formation
of spiral arms which reduce the strength of the small vortices in the main part of the disc.
On the other hand, the vortices created by the more massive planets, which were already visible in Fig.~\ref{fig:vortex}, are clearly identifiable as an increase in the vorticity. For the $\unit[100]{M_\oplus}$ model a vortex forms at the inner edge after $\unit[100]{orbits}$ at \unit[0.8]{$r_\mathrm{p}$} and starts to migrate inwards. The migration is initially slow, but it increases after leaving the gap edge. After further $\unit[100]{orbits}$ it migrates with a speed of approximately $0.001 r_\mathrm{p}$ per orbit to \unit[0.73]{$r_\mathrm{p}$} where it can be seen in the figure, which is on the same order of magnitude as found by \citetads{2013A&A...559A..30R}. Vortex migration has already been studied by \citetads{2010ApJ...725..146P} and is due to the angular momentum transport by the excited density waves, that are asymmetric due to the density and vorticity gradient, leading to a net torque and migration. The vortex at the outer edge does not migrate notably, because the presence of the planet prevents inward migration.

The $\unit[30]{M_\oplus}$ model has also an increased vorticity at the edges of the gap. While no large vortices have formed yet, 
two weak vortices are visible at the outer edge of the gap (see also Fig.~\ref{fig:vortex}).
This is in contrast to the $\alpha$-disc simulations, where for the $\unit[30]{M_\oplus}$ case no vortices are seen and only weak signatures can be noticed for the $\unit[100]{M_\oplus}$ case.
The last row shows the vorticity for the $\alpha$-disc, where small vortices are present in the horseshoe region for the $\unit[30]{M_\oplus}$ model, since the relatively small viscosity cannot dissipate them on a short timescale. The variation in vorticity in the outer region is due to VSI activity, which is not suppressed completely by the viscosity leaving mainly the modes with small wavenumber. Similar to the VSI disc, we see stronger VSI activity in the $\alpha$-disc in the $\unit[100]{M_\oplus}$ model both in the inner and outer disc.

\subsection{Surface density profile}
\begin{figure*}[t!]
    \centering
    \includegraphics{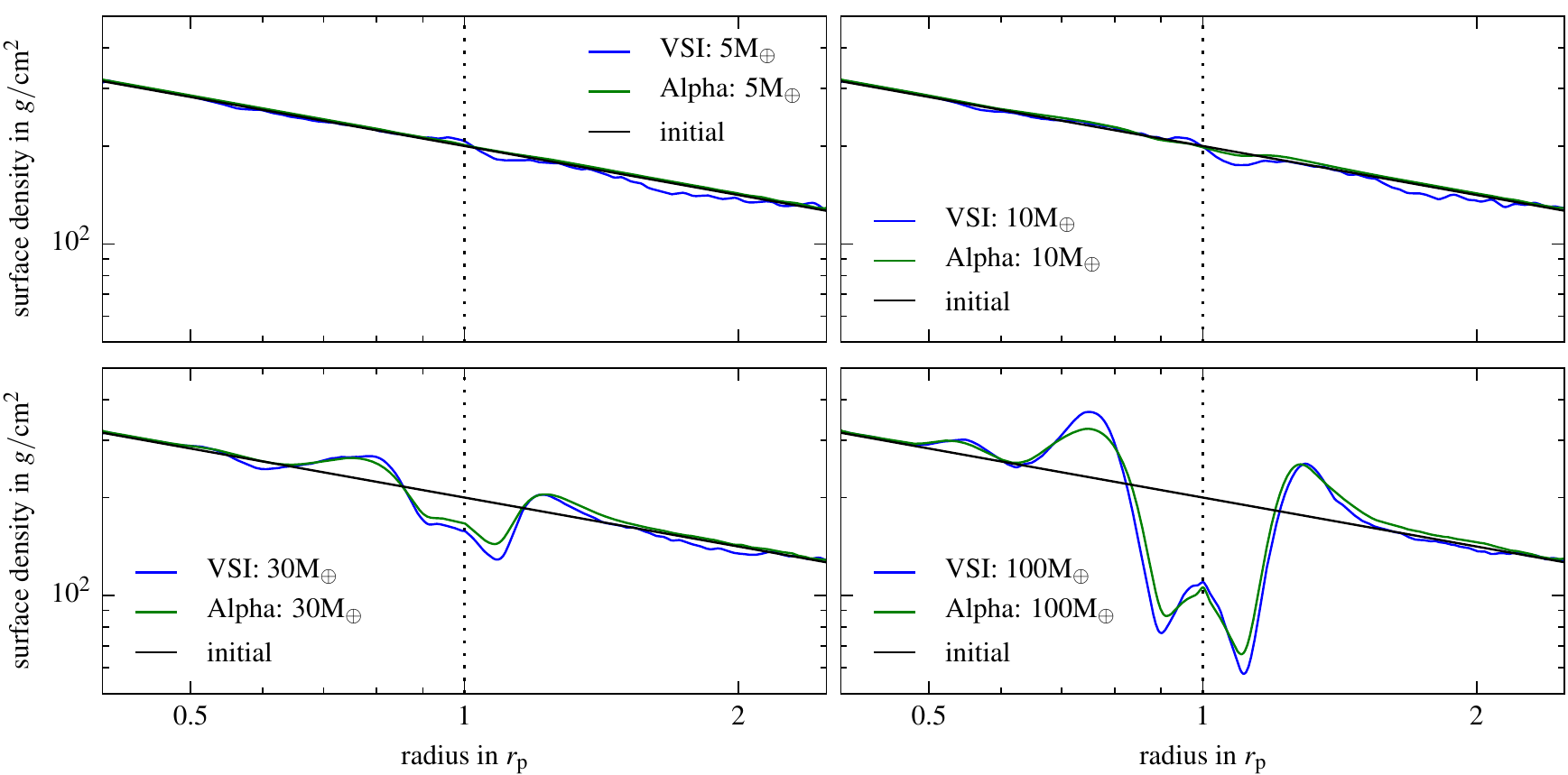}
    \caption{Surface density for the different simulations after $\unit[200]{orbits}$.
  The vertical dotted black lines indicate the position of the planet and the initial surface density profile is marked by the solid black line.
   For each planet mass the VSI models are compared to the $\alpha$-disc cases.}
    \label{fig:surface}
\end{figure*}

An important observable that can be studied is the surface density profile which, modified by the presence of the planet and local instabilities, can show specific recognisable patterns.
We present in Fig. \ref{fig:surface} the disc surface density distribution, with respect to the initial model, for the different models and planetary masses, averaged over the azimuthal direction.
 
In the VSI-disc the planet is able to carve slightly deeper gaps with respect to the $\alpha$-disc models. Moreover, the density is increased at the inner gap edge. These effects are caused by an enhanced vorticity near the gap edges, due to the interplay of the VSI with the Rossby wave instability \citepads[see][]{2016MNRAS.456.3571R}.
Surprisingly, even the smaller planets create a perturbation in the surface density profile close to the planet (especially in the VSI models). This is not a typical gap, but instead an increase in the surface density profile close the planet at the inner side of the disc and a decrease at the outer side. This comes from an increased vorticity in this region as we discussed in more detail in the Sec.~\ref{sec:vortex}.
Finally, an overall slight decrease in the surface density in the outer region is visible for all VSI models, partly because they have been evolved for $\unit[200]{orbits}$ before adding the planet, and also due to higher $\alpha$-parameter in the outer region (see Fig.~\ref{fig:alpha}). For the larger planets the surface density profiles show very good agreement between the VSI models and the corresponding viscous $\alpha$-disc models.
The gaps have the same widths and depths, confirming the estimate for the effective viscosity, $\alpha=5 \cdot 10^{-4}$.
%
%
\subsection{Torques acting on the planet}\label{sec:torques}
We calculate the torques using a smoothed force of the planet onto the disc
\begin{equation}
    \vec{F_\mathrm{p}} = \frac{G M_\mathrm{p} \Delta m}{d^3 + (\varepsilon R_\mathrm{Hill})^3} \vec{d} \,,
\end{equation}
where $d$ is the distance of a grid cell with mass $\Delta m$ from the planet and $\varepsilon = 0.5$ the smoothing parameter. The smoothing reduces the potential inside the Hill radius, where we get too large contributions otherwise from the high gas density due to the missing accretion of the gas onto the planet. Moreover the density distribution close to the planet should be symmetrical with respect to the planet and thus it should not contribute to the total torque, but the finite resolution can make this numerically challenging.

We compare our results to the simulations of \citetads{2010ApJ...724..730D} and use the same normalization for the total torque
\begin{equation}
    \Gamma_0 = \Sigma(r_\mathrm{p}) \Omega^2(r_\mathrm{p}) r_\mathrm{p}^4 \left( \frac{M_\mathrm{p}}{M_s} \right)^2 \left( \frac{r_\mathrm{p}}{H} \right)^2 \,,
    \label{norm}
\end{equation}
and for the torque distribution per unit disc mass as a function of radius
\begin{equation}
    \left( \frac{d\Gamma}{dm} \right)_0 =  \Omega^2(r_\mathrm{p}) r_\mathrm{p}^2 \left( \frac{M_\mathrm{p}}{M_s} \right)^2 \left( \frac{r_\mathrm{p}}{H} \right)^4 \,.
    \label{norm2}
\end{equation}

\begin{figure*}[t!]
    \centering
    \includegraphics{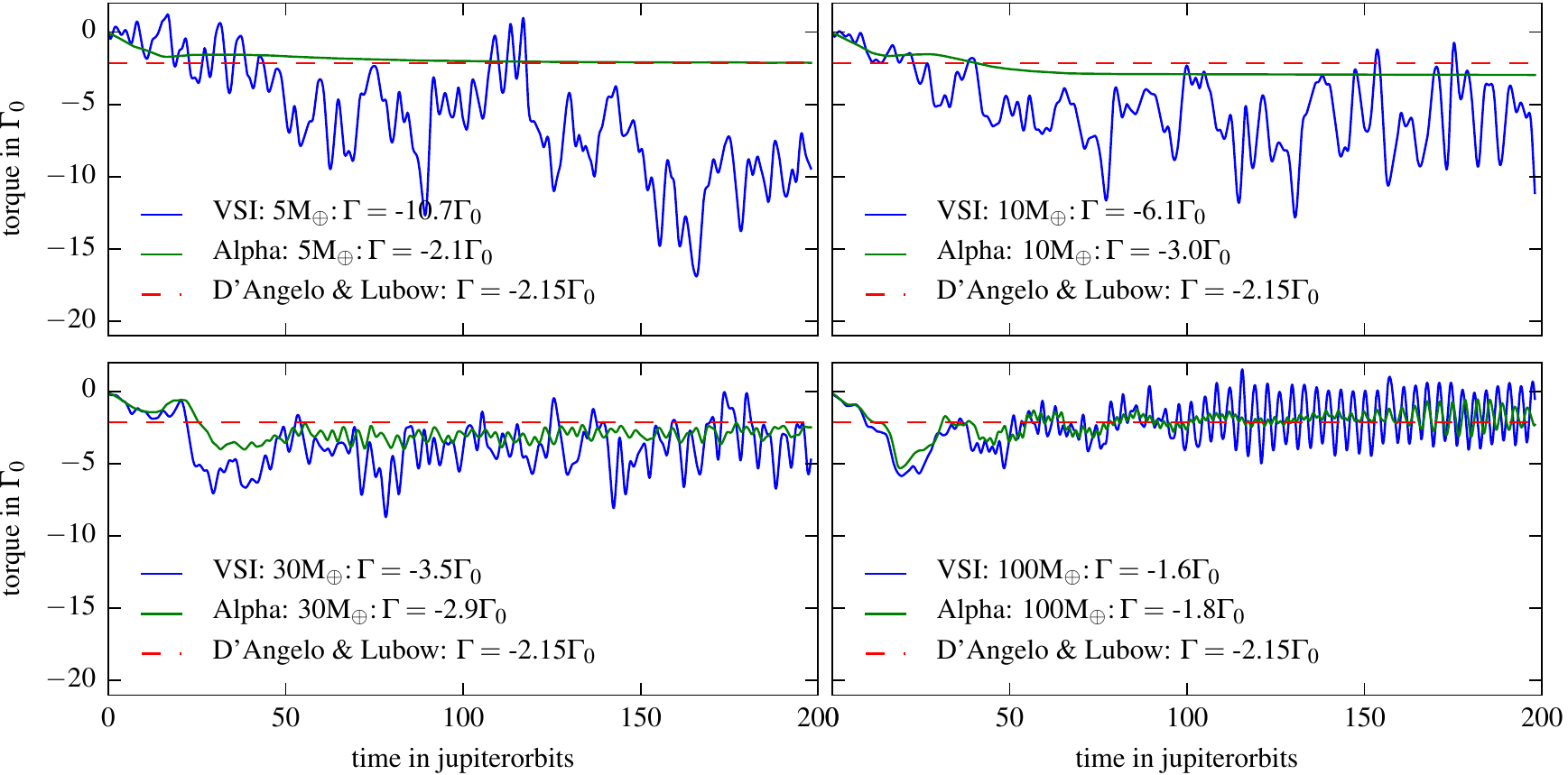}
    \caption{Torque acting on the planet over time normalised by $\Gamma_0$. We smooth over $\unit[4]{orbits}$ with a Gaussian window and compare to the simulation of \citetads{2010ApJ...724..730D} (red dashed line). In the legend we give the total torque averaged over the last $\unit[50]{orbits}$. Due to the turbulence in the VSI disc the torque strongly fluctuates while for the $\alpha$-disc the evolution is much smoother. The oscillations in the last panel for the $100 M_\oplus$ planet are due to the presence of a vortex.}
    \label{fig:torqueTime}
\end{figure*}

We display the torques for the VSI and $\alpha$-disc models over time in Fig. \ref{fig:torqueTime}, where we quote the total torque averaged over the last $\unit[50]{orbits}$ in the legend. We smoothed the torques over $\unit[4]{orbits}$ with a Gaussian window, to reduce the strong fluctuations. The source of the perturbations are different depending on the size of the planet. For the smaller planets they are introduced through the turbulence injected by the VSI. In the case of the $100 M_\oplus$ planet the more regular oscillations are introduced 
by a vortex near the outer edge of the gap generated by the planet.
From the last panel in Fig.~\ref{fig:torqueTime} we can estimate an oscillation period in this case of about 3.5 $P_\mathrm{p}$ 
which translates into a position of the vortex at $r \approx 1.25 R_\mathrm{p}$, in good agreement with Fig.~\ref{fig:vortex}.

We compare the torques to the simulations of \citetads{2010ApJ...724..730D}, who obtained for locally isothermal 3D discs the following empirical function
\begin{equation}
    \Gamma = - (1.36 +0.62 \beta + 0.43 \zeta) \Gamma_0 \,,
    \label{Lubow}
\end{equation}
where $\beta = - d\ln{\Sigma}/d \ln{r} $ and $\zeta = -d \ln{T} / d \ln{r}$, which correspond for our simulations to $\beta = 0.5$, $\zeta = 1$,
and lead to an expected torque of $\Gamma =-2.15 \Gamma_0$.

The torques on the planets in the viscous disc with $\alpha$-viscosity agree closely with the empirical function,
with small differences caused possibly by the different smoothing methods of the planetary potential adopted.
Only the $\unit[100]{M_\oplus}$ planet has a reduced torque due to gap formation.

In contrast, the simulations with VSI strongly disagree with the predictions. While it is not very surprising that the torques can be positive for a few orbits in the context of turbulent discs, it is clearly visible that the average of the torque is far from the predicted value. In case of the smallest planet with five Earth masses the average of the absolute torques is even five times larger than the torques in the $\alpha$-disc,
leading to a very rapid inward migration.
This effect becomes smaller with increasing planet mass 
because the larger planets begin to open gaps and generate larger vortices that are able to quench the VSI modes.

\begin{figure}[tb]
    \centering
    \includegraphics{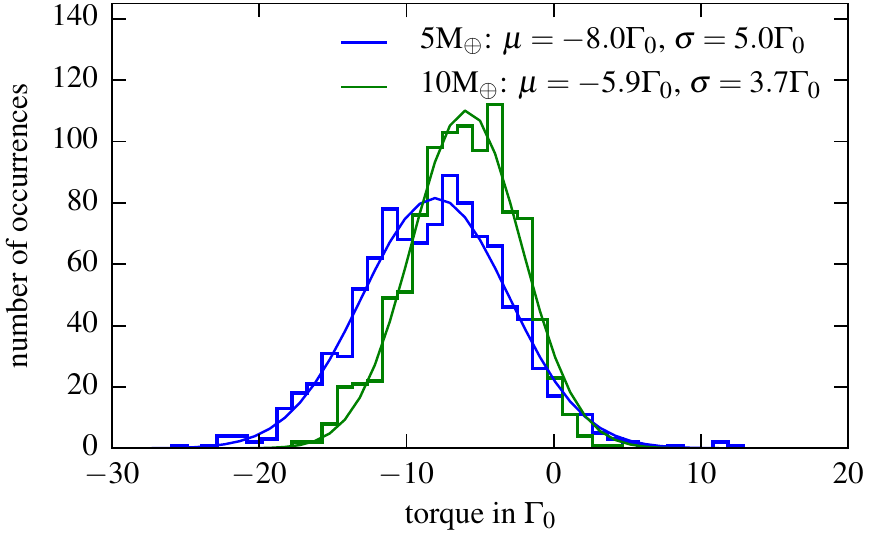}
    \caption{A histogram, taken over 1000 samples of the occurrence of certain torque values for the two low mass
     planets in the VSI turbulent disc. The data are collected for the last 100 orbits of the simulations.
    The torques are normalized to $\Gamma_0$ in each case
    and the average values and standard deviation over this time span are quoted in legend.}
    \label{fig:torque-histo}
\end{figure}

The time to reach an equilibrium torque is about 50 orbits for the $10 M_\oplus$ planet and somewhat longer for the smallest one.
To further analyse the stochastic nature of the perturbation on the planets and the resulting torque we display 
in Fig.~\ref{fig:torque-histo} the occurrence rate of torque values for the two low mass planets. 
As shown, they are approximately shaped like a Gaussian and centered around their mean values that
are quoted in the legend.
The standard deviations derived from fitting the Gaussians are 5.0 and 3.7 for the two low mass planets, such that the
strength of the torque fluctuations is comparable to the mean torque for both cases.
This is different to simulations of embedded planets in MHD turbulent disc. In the simulations performed for example by
\citet{2005A&A...443.1067N} and \citet{2011ApJ...736...85U} it is shown that the stochastic part of the torque can exceed the mean value
significantly and the timescale for reaching equilibrium is also longer in MHD turbulence. The reason for this difference lies
in the fact that the density fluctuations are stronger in the MHD turbulence compared to the VSI case.

\begin{figure}[tb]
    \centering
    \includegraphics{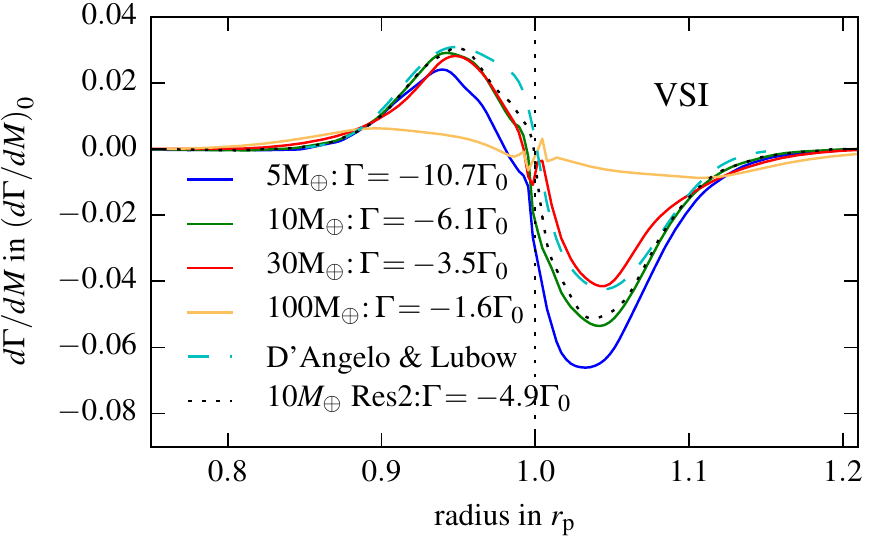}
    \includegraphics{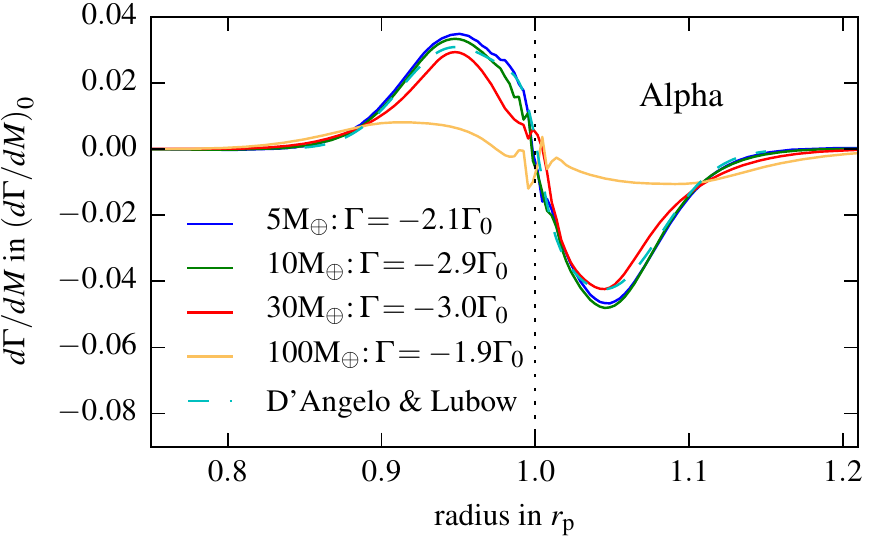}
    \caption{Torques on the planets normalised by $(d\Gamma/dM)_0$, see eq.~(\ref{norm2}). {\bf Upper Panel:} The planet is embedded into a disc with VSI. The inner torques are smaller then expected and the outer torques are stronger. {\bf Lower Panel:} The planet is embedded into a disc with generic $\alpha$-viscosity. The torques are very similar to the results of \citetads{2010ApJ...724..730D}. In both simulations the torques for the heaviest planet are reduced due to the lower surface density in the gap.}
    \label{fig:torqueRadial}
\end{figure}

\begin{figure}[tb]
    \centering
    \includegraphics{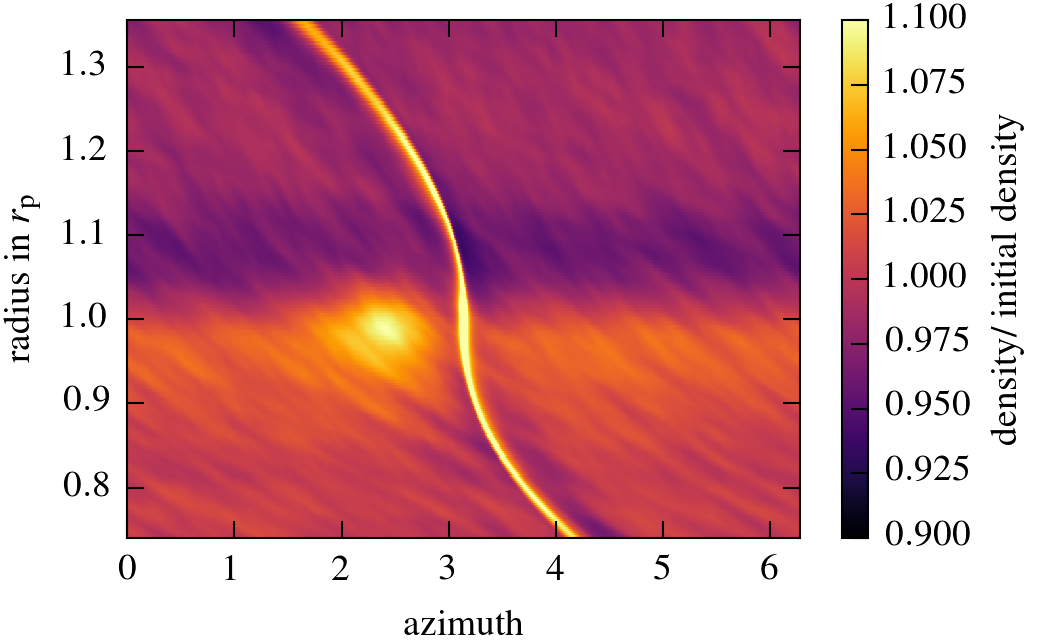}
    \includegraphics{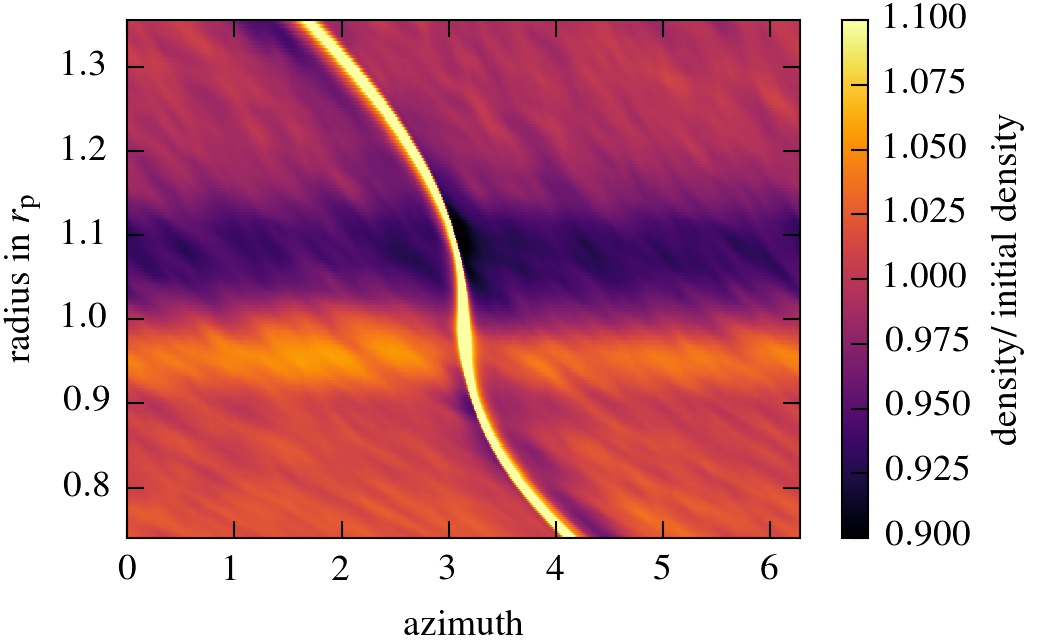}
    \caption{{\bf Upper Panel}: Density averaged over the last $\unit[100]{orbits}$ for the VSI simulation with $M_\mathrm{p}=\unit[5]{M_\oplus}$ in units of initial density. {\bf Lower Panel}: The density averaged over the last $\unit[100]{orbits}$ for the VSI simulation with $M_\mathrm{p}=\unit[10]{M_\oplus}$ in units of initial density. In both cases an increase in density before the planet is visible, if only barely for the $\unit[10]{M_\oplus}$ case. The bump in density does not move over time.}
    \label{fig:densityAverage}
\end{figure}

To shed more light on this torque increase for the low mass planets we also show the torque distribution per unit disc mass as a function of radius in Fig. \ref{fig:torqueRadial}. Here we average again over the last $\unit[50]{orbits}$ and we compare to \citetads{2008ApJ...685..560D} who give an analytical fit to the radial torque distribution.
In the lower panel of Fig. \ref{fig:torqueRadial} we can again see a good agreement with their results for the viscous disc, except for the heaviest planet.  In the simulation with $M_\mathrm{p} = \unit[5]{M_\oplus}$ the torque is five times stronger with active VSI compared to the simulation with $\alpha$-viscosity. This is due to a combination of reduced outer torque and stronger inner torque. 
This cannot be  explained by a variation in the radial surface density, because we could see in Fig. \ref{fig:surface} that the density in the outer region is reduced and the density in the inner region is enhanced, leading to the opposite effect. Instead the surface density depends on the azimuthal direction, and there we find an increase in density on the inner side behind the planet and a decrease in density at the outer side in front of the planet. This can be seen in Fig. \ref{fig:densityAverage} where we average the density over $\unit[100]{orbits}$. The density enhancement is due to the increased vorticity behind the planet, resulting from the interplay between the planet and the VSI as can be seen in the upper panels of Fig.~\ref{fig:vorticity2D}.

\subsection{A model with double resolution}
\label{subsec:double}
To check the convergence with resolution, we repeat the simulation for $M_\mathrm{p}=\unit[10]{M_\oplus}$ and VSI with doubled resolution in every direction. The resolution is then $1200\times256\times2048$ in (r, $\theta$, $\varphi$) or (15,11,7) cells per $R_\mathrm{Hill}$ for this planet. The overall results show no important difference to the results of the simulations with lower resolution. A notable difference is in the $\alpha$-profile, which we present in Fig. \ref{fig:alpha}. 
We can see that in the inner region the $\alpha$-parameter is larger for the better resolved simulation. 
This indicates that the simulations using the standard resolution are not properly resolved in the innermost region between \unit[0.5-0.64]{$r_\mathrm{p}$},
which could be a consequence of the fact that the radial extent of the VSI cells increases stronger than linear with radius,
as shown in \citet{2014A&A...572A..77S}. 
This also indicates that the rest of the domain is sufficiently resolved and not dependent on resolution, in contrast to the 2D simulations in \citetads{2014A&A...572A..77S}. This does not pose a problem for the calculations of the torques, which are only important close to the planet. This can be seen in Fig. \ref{fig:torqueRadial}, were the dotted black line represents the simulation with doubled resolution which can be compared to the green line. Both lines are as close together as one can expect from simulations with turbulence. No other notable differences have been observed.

\section{Discussion and Conclusions} \label{sec:discussions}
We performed a series of hydrodynamical simulations to study the interplay between a growing planet embedded in a protoplanetary disc and the VSI, which is a strong candidate for the main angular momentum transport mechanism in the region of active planet formation. The main results that can be drawn from our study are:
\begin{enumerate}
\item Vortices and VSI \\
Larger planetary cores ($M_\textrm{p} > \unit[10]{M_\oplus}$) are able to open easily a gap in the disc, which leads to the generation of vortices at the gap edges.
These vortices are observed both in VSI and $\alpha$-disc models, however, due to the interaction between the vortices and the VSI,
their lifetime is much longer in the VSI models.
In our simulations we notice the presence of several weak spiral arms, for example in Fig.~\ref{fig:vortex} (upper row), that can be associated with vortices generated by the Rossby-Wave Instability (RWI) near the edges of the gap created by the planet. They are related to a maximum in the normalised vorticity profile, shown in Fig.~\ref{fig:vorticity2D}. Furthermore, the vorticity perturbations generated by the VSI, as suggested by \citetads{2016MNRAS.456.3571R}, are effectively increasing the lifetime of the RWI vortices, as one can see comparing the viscous and turbulent runs in the bottom panel of Fig.~\ref{fig:vorticity2D}.

\item {The VSI active region} \\
The region with active VSI is not affected sensibly by the embedded planets for planetary masses less than $\unit[30]{M_\oplus}$. 
For higher masses the presence of vortices is able to  suppress the VSI near their position.
On the other hand the active region is extended inwards, at least for the most massive planet studied (see Fig. \ref{fig:vorticity2D}).
One possibility for this extended turbulent activity might by the action of the spiral wave instability (SWI)
where the presence of a massive planet can excite unstable inertial-gravity waves \citep{2016ApJ...829...13B,2016ApJ...833..126B}
which might enhance the turbulent level above that of the VSI alone.
We have analysed the time evolution of the vertical kinetic energy and 2D slices of the vertical velocity but could not find a
clear indications for the operation of the SWI.
A possible other explanation for this effect is the change in surface density profile due to the creation of a gap.
This will in turn change the angular momentum profile and thus influence the VSI.
\item {Planet migration} \\
Concerning the migration of planets we find that
the torques acting onto small planets for the $\alpha$-disc models are in good agreement with the prediction of
\citet{2010ApJ...724..730D}. This picture changes drastically for the inviscid models with the VSI operating.
For small planets with masses smaller than about $\unit[10]{M_\oplus}$ the time averaged torques are negative throughout.
For the lowest planet mass we studied, $\unit[5]{M_\oplus}$, the inward migration is about 5 times faster
than for the $\alpha$-disc model which is due to a density enhancement directly behind the planet.
For increasing planet masses the migration rates approach those of the viscous disc. Due to this special disturbance in the density
we do not find any prolonged phases of outward migration as seen for example for planets embedded in MHD-turbulent discs
\citep{2005A&A...443.1067N,2011ApJ...736...85U}. In general, because the density fluctuations in VSI turbulent discs are smaller
than in MHD turbulence, the stochastic component in the migration process is weaker in VSI discs.
Even though the stochastic component does not play a major role for our small planet case with $\unit[5]{M_\oplus}$,
it may nevertheless become important for sub-Earth mass objects.
As three-dimensional inviscid simulations of disc will always generate turbulence through the VSI, embedded planets
will experience substantial inward migration. This feature cannot be captured by modelling planet migration in two-dimensional
inviscid discs that show typically a stalling of migration after a sufficiently wide gap has been formed \citep{2009ApJ...690L..52L,2017ApJ...839..100F}.
In contrast, realistic three-dimensional discs will always show a finite turbulence level that will limit the gap depth.
In fact, from our simulations we see that the width and depths of gaps opened by the planets are very similar in the VSI models and the
corresponding $\alpha$-disc models.
In addition, small planets ($M_\textrm{p} \leq \unit[10]{M_\oplus}$) that do not open significant gaps are able to modify the surface density profile
close to their location which leads to the enhanced migration for the VSI turbulent discs.

\item {Vortex migration and its implications} \\
Vortex migration was first observed by \citetads{2010ApJ...725..146P} in isothermal discs, and then further studied by \citetads{2015A&A...573A.132F} for radiative discs. Vortices are able to generate spiral arms by compressing the flow around them, and the migration occurs because of an asymmetry in the position of the sonic lines that generates the density waves. According to \citetads{2013A&A...559A..30R} the migration speed is directly linked with the vortex aspect ratio ($\chi$), it slows down upon increasing $\chi$. In particular, \citetads{2016A&A...586A.105F} studied the interaction of a planet with a vortex generated through RWI.
Interestingly, they found that a massive vortex can drag the planet with it during the migration process, and it can also cross the planetary gap periodically crossing its location.

In our simulations for the $\unit[100]{M_\oplus}$ planet, the inner vortex has a migration rate after an initial adjustment on the order of $\unit[0.001]{r_\mathrm{p}/orbit}$, in accordance with \citetads{2013A&A...559A..30R}, while the outer one is kept in its orbit by the steep surface density profile carved by the planet.
The aspect ratio of the inner vortex is around $3$ which, according to \citetads{2013A&A...559A..30R} should be destroyed after a few orbital periods due to elliptical instability (for $\chi<4$). However, its lifetime is much more extended in our runs, meaning that the presence of the VSI is beneficial extending its life. On the other hand the outer vortex is much broader and has a larger aspect ratio.
Both vortices have also a considerable vertical extent, and only in the outer corona they are effectively dissipated.

\item {VSI as an angular momentum driving process} \\
In agreement with previous simulations of VSI unstable discs \citetads{2013MNRAS.435.2610N},
we find that an angular momentum transport driven by the VSI that corresponds to an $\alpha = 5\cdot10^{-4}$ which is not strongly
affected by embedded planets. Hence, the VSI constitutes a viable candidate for the generation of turbulence in discs where the MRI may be
inactive. Recently, we have shown \citep{2017A&A...599L...6S} that the turbulence generated by the VSI is anisotropic and can be described
by a viscous ansatz for the stress tensor using two different coefficients for the radial and vertical angular momentum transport.
It remains to be seen how an embedded planet behaves in such discs with anisotropic viscosities.
\end{enumerate}

Thus, the VSI can on the one hand increase the strength of the vortices forming close to the gap edges for high planetary masses, and on the other hand boost the Type I migration for small planetary cores. 
The present study can be further improved by relaxing the locally isothermal assumption, and 
perform radiative simulations over a similar radial range to check the strength of the VSI turbulence and its impact
on embedded planets under more realistic conditions. Additionally, in further improvements the planet should be 
allowed to accrete mass and migrate through disc.

\begin{acknowledgements}
    We thank an anonymous referee for useful comments and suggestions.
    G. Picogna acknowledges the support through the German Research Foundation
    (''Deutsche Forschungsgemeinschaft'', DFG) grant KL 650/21 within the 
    collaborative research program {\it The first
    10 Million Years of the Solar System}. M.H.R. Stoll
    acknowledges the support through the (DFG) grant KL 650/16.
    This work was performed on the computational resource ForHLR I funded by
    the Ministry of Science, Research and the Arts of Baden-W\"urttemberg and the DFG. 
\end{acknowledgements}

\bibliographystyle{aa}
\bibliography{calibre,biblio,wk}{}



\end{document}